\documentstyle[aps,epsfig,twocolumn]{revtex}

\begin{document}
\title{Efficient engineering of multi-atom entanglement through single-photon
detections}
\author{L.-M. Duan$^{1,2}$, H. J. Kimble$^{3}$}
\address{$^{1}$Institute for Quantum Information, MC 107-81, California
Institute of Technology, Pasadena, CA 91125-8100\\
$^{2}$Laboratory of quantum information, USTC, Hefei 230026, China\\
$^{3}$Norman Bridge Laboratory of Physics 12-33, California
Institute of Technology, Pasadena, CA 91125} \maketitle

\begin{abstract}
We propose an efficient scheme to engineer multi-atom entanglement by
detecting cavity decay through single-photon detectors. In the special case
of two atoms, this scheme is much more efficient than previous probabilistic
schemes, and insensitive to randomness in the atom's position. More
generally, the scheme can be used to prepare arbitrary superpositions of
multi-atom Dicke states without the requirements of high-efficiency
detection and separate addressing of different atoms.

{\bf PACS numbers:} 03.67.-a, 42.50.-p, 42.50.Gy
\end{abstract}

There is a large current interest in generation and engineering of quantum
entanglement, with applications for fundamental tests of quantum mechanics
\cite{1}, for high-precision measurements \cite{2}, and in particular, for
implementation of quantum communication and computation \cite{3}. Although
quantum entanglement is typically fragile to practical noise and technical
imperfections, there exist elegant ways to overcome this sensitivity by
designing schemes with inherent robustness to diverse sources of noise. Some
schemes with this property have been known for entangling two single atoms
\cite{4,5,5',5'',6,7} as well as for entangling macroscopic atomic ensembles
\cite{8,9}. In these schemes, feedback is typically applied to the system of
interest based upon the outcome of certain measurements. The protocols are
thereby probabilistic, succeeding only conditionally for particular
measurement results. Imperfections and noise in these schemes decrease the
success probability, but have no influence on the fidelity of the intended
state generation for the \textquotedblleft successful\textquotedblright\
subset of trials. In this way, a high-fidelity entangled state can be
obtained simply by repeating the scheme successively.

Here, we propose a robust scheme to produce and engineer entanglement
between multiple atoms in optical cavities. Compared with the previous
robust schemes \cite{4,5,5',5'',6,7,8,9}, our protocol has the following
favorable features. {\it (i)} It is much more efficient in the sense that
the success probability can be close to unity, whereas in the previous
schemes \cite{4,5,6,7,8,9}, the success probability is required to be much
smaller than $1$ to have the property of inherent robustness. {\it (ii)} It
is more insensitive to certain practical sources of noise, such as
randomness in the atom's position, atomic spontaneous emission, or detector
inefficiency. {\it (iii)} Individual addressing of atoms is not required
\cite{5'}, nor are single photon states as initial resources \cite{5''}.{\it %
(iv)} Most importantly, our scheme is not limited to generation of two-atom
entanglement. Indeed, we show that based on current experimental technology,
it should be possible to generate any superposition of the Dicke states \cite
{9'} between multiple atoms in an optical cavity. These Dicke states and
their superpositions, including the multi-party GHZ states as special cases,
are typically highly entangled, and are useful for many applications in
quantum information science \cite{2,10,11,11'}.

As the scheme here is inherently robust to noise, it works in
principle for entangling atoms (or ions) both in free-space
configurations and in high-Q cavities, albeit in the free-space
case one has a much smaller success probability to collect the
emitted photons. In this paper, for a close relation with the
current experimental efforts \cite{12,12',12''}, we assume that
there is a standing-wave high-Q cavity around the atoms
\cite{13,13'} to improve the collection efficiency.

To explain the scheme, let us start from the simplest case with
two atoms trapped in two different cavities. The schematic setup
is shown in Fig. 1A, with the relevant atomic levels depicted in
Fig. 1B. The states $\left| g\right\rangle $, $\left|
0\right\rangle $, $\left| 1\right\rangle $ correspond to the
hyperfine and the Zeeman sublevels of alkali atoms in the
ground-state manifold, and $\left| e\right\rangle $ corresponds to
an excited state. The atom is initially prepared in the state
$\left| g\right\rangle $, but the basis-vectors of a qubit are
represented by the states $\left| 0\right\rangle $ and $\left|
1\right\rangle $. The transition $\left| g\right\rangle
\rightarrow \left| e\right\rangle $ is driven adiabatically
through a classical laser pulse with the corresponding Rabi
frequency denoted by $\Omega \left( t\right) $ \cite{14}. With the
driving pulse, the atom is transferred with probability
$p_{c}\simeq 1$ to the $\left| 0\right\rangle $ and $\left|
1\right\rangle $ states by emitting a photon from the transitions
$\left| e\right\rangle \rightarrow \left| 0\right\rangle $ or
$\left| e\right\rangle \rightarrow \left| 1\right\rangle $.
Without loss of generality, we assume that the transitions $\left|
e\right\rangle \rightarrow \left| 0\right\rangle $ and $\left|
e\right\rangle \rightarrow \left| 1\right\rangle $ are coupled to
two degenerate cavity modes $a^{h}$ and $a^{v}$ with different
polarizations $h$ and $v$. The decay pulses from the two cavities
are interfered at a polarization beam splitter (PBS), with the
outputs detected by two single-photon detectors after a $45^{o}$
polarizer (denoted as $P_{45}$ in Fig. 1A). The small fraction of
the transmitted classical pulse can be easily filtered based on
the frequency selection. For the decay pulse from the {\bf R}
cavity, a polarization rotator $R\left( \pi /2\right) $ is
inserted before the PBS which exchanges $h$ and $v$ polarizations
of the incoming photon. Conditioned upon registering one photon
from {\it each} of the detectors, the two atoms in the cavities
{\bf L} and {\bf R} will be prepared into the maximally entangled
state
\begin{equation}
\left| \Psi _{LR}\right\rangle =\left( \left| 01\right\rangle _{LR}+\left|
10\right\rangle _{LR}\right) /\sqrt{2}.
\end{equation}

To see this, we write down the interaction Hamiltonian in the rotating
frame, which, for each of the cavities, has the form (setting $\hbar =1$)
\begin{equation}
H=\Omega \left( t\right) \left| e\right\rangle \left\langle g\right|
+g_{0}\left| e\right\rangle \left\langle 0\right| a^{h}+g_{1}\left|
e\right\rangle \left\langle 1\right| a^{v}+H.c.,
\end{equation}
where $g_{0}$ and $g_{1}$ are the corresponding coupling rates. The cavity
outputs $a_{out}^{\mu }$ $\left( \mu =h,v\right) $ are connected with the
cavity modes $a^{\mu }$ through the standard input-output relations $%
\stackrel{.}{a}^{\mu }=-i\left[ a^{\mu },H\right] -\kappa a^{\mu }/2-\sqrt{%
\kappa }a_{in}^{\mu }\left( t\right) $ and $a_{out}^{\mu }\left( t\right)
=a_{in}^{\mu }\left( t\right) +\sqrt{\kappa }a^{\mu }$ \cite{15}, where $%
\kappa $ is the cavity decay rate, and $a_{in}^{\mu }\left( t\right) $, with
the commutation relation $\left[ a_{in}^{\mu }\left( t\right) ,a_{in}^{\mu
\dagger }\left( t^{\prime }\right) \right] =\delta \left( t-t^{\prime
}\right) $, denotes the vacuum cavity input. We are interested in the limit
for which the variation rate of $\Omega \left( t\right) $ is significantly
smaller than the cavity decay rate $\kappa $. In this limit, we can define
an effective single-mode bosonic operator $a_{eff}^{\mu }$ from the cavity
output operator $a_{out}^{\mu }\left( t\right) $ as $a_{eff}^{\mu
}=\int_{0}^{T}f\left( t\right) a_{out}^{\mu }\left( t\right) dt$ (see Refs.
\cite{15a,14}), where $T$ is the pulse duration and $f\left( t\right) $ is
the output pulse shape, which is determined by the shape of $\Omega \left(
t\right) $ as $f\left( t\right) =\sqrt{\kappa }\sin \theta \left( t\right)
\exp \left[ -\left( \kappa /2\right) \int_{0}^{t}\sin ^{2}\theta \left( \tau
\right) d\tau \right] $ with $\sin \theta \left( t\right) =\Omega \left(
t\right) /\sqrt{\left| g_{0}\right| ^{2}+\left| g_{1}\right| ^{2}+\left|
\Omega \left( t\right) \right| ^{2}}$. After the driving pulse, for each of
the cavities $\lambda $ ($\lambda =${\bf L},{\bf \ R}), the final state
between the atom and the corresponding cavity output has the form
\begin{equation}
\left| \Psi \right\rangle _{\lambda }=\left( g_{0}\left| 0\right\rangle
_{\lambda }\left| h\right\rangle _{\lambda }+g_{1}\left| 1\right\rangle
_{\lambda }\left| v\right\rangle _{\lambda }\right) /\sqrt{\left|
g_{0}\right| ^{2}+\left| g_{1}\right| ^{2}},
\end{equation}
where $\left| \mu \right\rangle =a_{eff}^{\mu \dagger }\left|
vac\right\rangle $, $\left( \mu =h,v\right) $, and $\left| vac\right\rangle $
denotes the vacuum state of the optical modes.

If the driving pulses have the same shape $\Omega \left( t\right) $ for the
{\bf L} and {\bf R} cavities, the output single-photon pulses from the two
cavities will also have the same shape $f\left( t\right) $, and they will
interfere with high visibility at the polarization beam splitter (PBS). If
one gets a ``click''\ from each of the detectors at the outputs of the PBS,
the two incoming photons can be either both in $h$ polarizations or both in $%
v$ polarizations, and these two possibility amplitudes are coherently
superposed when the incoming photon pulses overlap with each other with the
same shape. Therefore, the measurement in Fig. 1A, together with the
polarization rotator $R\left( \pi /2\right) $, corresponds to projecting the
whole state $\left| \Psi \right\rangle _{L}\otimes \left| \Psi \right\rangle
_{R}$ between the atoms and the photons onto a subspace with the projection
operator given by $P_{s}=\left| hv\right\rangle _{LR}\left\langle hv\right|
+\left| vh\right\rangle _{LR}\left\langle vh\right| $. Within this
measurement scheme, the state $\left| \Psi \right\rangle _{L}\otimes \left|
\Psi \right\rangle _{R}$ is effectively equivalent to the four-particle GHZ
state
\begin{eqnarray}
\left| \Psi _{eff}\right\rangle  &\propto &P_{s}\left| \Psi \right\rangle
_{L}\otimes \left| \Psi \right\rangle _{R} \\
&\propto &\left( \left| 01\right\rangle _{LR}\otimes \left| hv\right\rangle
_{LR}+\left| 10\right\rangle _{LR}\otimes \left| vh\right\rangle
_{LR}\right) /\sqrt{2}.  \nonumber
\end{eqnarray}
The $45^{o}$ polarizers in Fig. 1A project the photon polarizations to the $%
\left( \left| h\right\rangle +\left| v\right\rangle \right) /\sqrt{2}$
state. It immediately follows from Eq. (4) that after this measurement the
two atoms will be prepared in the maximally entangled state (1). If one
rotates the angles of the polarizers in Fig. 1A, corresponding a measurement
of the incoming photon polarizations either in the $\left\{ \left|
h\right\rangle ,\left| v\right\rangle \right\} $ basis or in the $\left\{
\left( \left| h\right\rangle +\left| v\right\rangle \right) /\sqrt{2},\left(
\left| h\right\rangle -\left| v\right\rangle \right) /\sqrt{2}\right\} $
bases, one can further demonstrate four-particle GHZ-type of entanglement
between the atoms and the photons as indicated by the effective state (4)
\cite{16}. The $45^{o}$ polarizer can also be replaced by a PBS with both of
its outputs detected by single-photon detectors. The measurement success
probability is then increased by a factor of $2$ for each side, and the
overall success probability of this scheme becomes $p_{s}=2\left|
g_{0}g_{1}\right| ^{2}/\left( \left| g_{0}\right| ^{2}+\left| g_{1}\right|
^{2}\right) ^{2}$.

Before introducing the multi-atom entangling scheme, we offer a few remarks
about this two-cavity scheme. First, it is evident that the scheme is
inherently robust to atomic spontaneous emission, output coupling
inefficiency, and detector inefficiency, all of which contribute to loss of
photons. Since a ``click''\ from each of the detectors is never recorded if
one photon is lost, these processes simply decrease the success probability $%
p_{s}$ by a factor of $\eta ^{2}$ (where $1-\eta $ denotes the loss for each
of the photons), but have no influence on the fidelity of the final state $%
\left| \Psi _{LR}\right\rangle $. Second, our scheme does not require
localization of the atom in the cavity to the Lamb-Dick limit. For the
standing-wave cavity shown in Fig. 1A and with the collinear pumping
configuration proposed in Ref. \cite{14}, $\Omega \left( t\right) ,$ $g_{0},$
and $g_{1}$ depend on the atom's position through approximately the same
cavity mode function. The pulse shape $f\left( t\right) $, which is
determined by the ratios $\Omega \left( t\right) /g_{0}$ and $\Omega \left(
t\right) /g_{1}$, thus becomes basically independent of the random variation
in the atom's position. For a travelling-wave cavity or for a free-space
configuration, the atom's position only affects the common phase of the
coupling rates $g_{0}$ and $g_{1}$, and in this case, a transverse pumping
configuration also suffices since the randomness in the common phase of $%
g_{0}$ and $g_{1}$ has no influence on the final entangled state $\left|
\Psi _{LR}\right\rangle $. Finally, the success probability of our scheme is
$p_{s}\sim 1/2$ in the ideal case with $g_{0}\sim g_{1}$ and $\eta \sim 1$,
which shows that the present scheme is significantly more efficient than the
previous schemes \cite{4,5,6,7,8,9}, where the success probability is
required to be much smaller than $1$ even if $\eta \rightarrow 1$.

We next extend our basic scheme to entangle multiple atoms in the same
optical cavity. The schematic setup is shown by Fig. 2, with each of the $%
N_{a}$ atoms taken to have the same level structure as depicted in
Fig. 1B and with the atoms not separately addressable
\cite{12,12',12''}. The initial state of the system has the form
$\left| G\right\rangle =\bigotimes_{i=1}^{Na}\left| g\right\rangle
_{i}$ with all the atoms prepared to the ancillary state $\left|
g\right\rangle $. The driving laser, incident from one side
mirror, is now divided into $M$ sequential pulses, with $M\geq
N_{a}/2$. We assume that the intensity of the pulse is controlled
so that for each of the $M$ pulses, an approximate fraction $1/M$
of the atomic population is transferred adiabatically from the
$\left| g\right\rangle $ state to the the $\left| 0\right\rangle $
or $\left| 1\right\rangle $ states, by emitting on average
$N_{a}/M$ photons with $h$ or $v$ polarizations. The output
photons from the cavity decay are split by a PBS according to
their polarizations, and then registered through two single photon
detectors (called $h$ and $v$ detectors, respectively). For each
driving pulse, we may or may not get a click from the $h$ or $v$
detectors, which are assumed {\it not} to distinguish one or more
photons.
For the whole $M$ pulses, we can count the total number of ``clicks''\ $%
(n_{h},n_{v})$ registered from the $(h,v)$ detectors, respectively. Of
course, $n_{h}+n_{v}\leq N_{a}$ since there are only $N_{a}$ atoms. If it
turns out that $n_{h}+n_{v}=N_{a}$, the following Dicke state results for
the $N_{a}$ atoms:
\begin{equation}
\left| N_{a},n_{h}\right\rangle =c\left( n_{h}\right) \left( s_{0}^{\dagger
}\right) ^{n_{h}}\left( s_{1}^{\dagger }\right) ^{N_{a}-n_{h}}\left|
G\right\rangle \text{.}  \label{dicke}
\end{equation}
Here, the collective operators $s_{\mu }^{\dagger }$ $\left( \mu =0,1\right)
$ are defined as $s_{\mu }^{\dagger }=\sum_{i=1}^{N_{a}}\left| \mu
\right\rangle _{i}\left\langle g\right| $, and the normalization coefficient
$c\left( n_{h}\right) =n_{h}^{-1}\left( N_{a}-n_{h}\right) ^{-1}\left[
N_{a}!/(n_{h}!\left( N_{a}-n_{h}\right) !)\right] ^{-1/2}$ if{\it \ }$%
n_{h}\neq 0,N_{a}$ ($c\left( 0\right) =c\left( N_{a}\right) =1/N_{a}$).
Except the trivial cases with $n_{h}=0,N_{a}$, clearly the Dicke state $%
\left| N_{a},n_{h}\right\rangle $ is entangled. The multi-atom Dicke states
and the GHZ\ states in general belong to different classes of entangled
states, and the Dicke states are relatively more robust to the influence of
noise \cite{10}. The Dicke states have some interesting applications in
quantum information processing and in high precision measurements \cite
{11,11'}.

To understand why a Dicke state results conditioned upon the above type of
measurement, we note that each atom has an equal probability to emit a
photon with the same pulse shape for each driving pulse for the assumed
sequence of adiabatic passages. Hence, each driving pulse involves a
collective excitation of the atoms to the $\left| 0\right\rangle $ or $%
\left| 1\right\rangle $ levels with homogeneous superposition coefficients.
For the subset of measurements for which we register $N_{a}$ photoelectric
events in total from the $h$ and $v$ detectors for the whole $M$ pulses,
each ``click''\ of the detectors should correspond exactly to the emission
of one and only one photon by the atoms. This is the case even if there are
photon loss and detector inefficiencies, because we post select only the
trials with exactly $N_{a}$ photoelectric events. Therefore, for each
``click''\ registered from the $h$ or $v$ detectors for these trials, we
should apply\ correspondingly the collective operators $s_{0}^{\dagger }$ or
$s_{1}^{\dagger }$ to the atomic state. After registering $n_{h}$ $h$%
-polarized photons and $\left( N_{a}-n_{h}\right) $ $v$-polarized photons,
we get exactly the state of Eq. (\ref{dicke}). Similar to the two-cavity
scheme, this multi-atom entangling scheme is also robust to practical
imperfections, such as a moderate randomness in the atoms' positions and
various sources of photon loss. Again, photon loss reduces the success
probability instead of the state fidelity.

To calculate the success probability of the multi-atom entangling scheme, we
note that the stepwise driving method described above is actually equivalent
to the following one-step driving method: we transfer all the atomic
population to the $\left| 0\right\rangle $ and $\left| 1\right\rangle $
levels with a single driving pulse, but both of the $h$ and $v$ polarized
photons after the PBS need to be further split equally into $M$ paths
through a series of beam splitters, with separate photoelectric detection
for each path. The state in Eq. (\ref{dicke}) corresponds to the case when $%
n_{h}$ $h$-detectors and $\left( N_{a}-n_{h}\right) $ $v$-detectors register
a photoelectric event. When two or more photons go to the same path, the
number of detector events is certainly less than $N_{a}$. So, for overall
success with $N_{a}$ events, we require that each photon follows a distinct
path, for which the success probability is given by $p_{si}=\left( 2M\right)
!/\left[ \left( 2M-N_{a}\right) !\left( 2M\right) ^{N_{a}}\right] $ (in
total there are $2M$ paths. For simplicity, we have assumed $g_{0}\approx
g_{1}$ so that one has equal probability to get $h$ or $v$ photons.) All
photon loss processes simply contribute to an under-count probability $%
1-\eta $ for each photon. Hence, the success probability to generate one of
the Dicke states of Eq. (\ref{dicke}) is $p_{succ}=\eta ^{N_{a}}p_{si}$,
while the probability to obtain a specific Dicke state $\left|
N_{a},n_{h}\right\rangle $ is $p_{n_{h}}=p_{succ}2^{-N_{a}}N_{a}!/\left[
n_{h}!\left( N_{a}-n_{h}\right) !\right] $. Excluding the trivial cases with
$n_{h}=0,N_{a}$, we then find that the success probability to obtain an
entangled state from this scheme is $p_{en}=p_{succ}\left(
1-2^{-N_{a}+1}\right) $, which tends to unity in the case $2M>>N_{a}$ if we
neglect contributions from photon loss (i.e., $\eta \rightarrow 1$). This
scheme could thus be quite efficient. For instance, with $\eta =0.70$ and $%
M=50$ pulses, $p_{en}=0.018$ for $N_{a}=10$ atoms, so that repeating this
scheme on average $1/p_{en}\approx 56$ times leads to a high-fidelity
entangled state between $10$ atoms. In current experimental setups \cite
{12,12'}, the typical duration $\Delta t$ of the adiabatic pulse is a few
hundred nanoseconds, so that the total duration $\left( M/p_{en}\right)
\Delta t\simeq 10^{-3}$ s.

Finally, we would like to extend further the above scheme to generate any
superposition of the Dicke states $\left| N_{a},n_{h}\right\rangle $. For
this purpose, we simply insert a polarization rotator $R\left( \theta
,\varphi \right) $ before the PBS as shown in Fig. 2, which transforms the
photon polarizations according to $\left| h\right\rangle \rightarrow \cos
\theta \left| h\right\rangle +\sin \theta e^{i\varphi }\left| v\right\rangle
$ and $\left| v\right\rangle \rightarrow -\sin \theta e^{-i\varphi }\left|
h\right\rangle +\cos \theta \left| v\right\rangle $. We assume that the
parameters $\theta ,\varphi $ can be separately controlled for each driving
pulse, and are denoted by $\theta _{m},\varphi _{m}$ for the $m$th pulse. As
before, we consider only the subset of cases for which exactly $N_{a}$
photoelectric events are registered from the whole $M$-pulse sequence. If
the $h$ (or $v$) detector ``clicks''\ for the $m$th pulse with the control
parameters $\theta _{m},\varphi _{m}$, the corresponding atomic excitation
operator $P_{m0}$ (or $P_{m1}$) is expressed by the collective operators $%
s_{\mu }^{\dagger }$ as $P_{m\mu }=\cos \theta _{m}s_{\mu }^{\dagger
}-\left( -1\right) ^{\mu }\sin \theta _{m}e^{i\varphi _{m}\left( -1\right)
^{\mu }}s_{1-\mu }^{\dagger }$ ($\mu =0,1$). So, after $N_{a}$ registered
events, the final atomic state has the form $\left| \Psi _{F}\right\rangle
=\prod_{i=1}^{N_{a}}P_{m_{i}\mu }\left| G\right\rangle $, where $m_{i}$ $%
\left( i=1,2,\cdots ,N_{a}\right) $ denote the set of driving pulses for
which we register a photon. Each operator $P_{m_{i}\mu }$ introduces two
real parameters $\theta _{m_{i}},\varphi _{m_{i}}$, so there are $2N_{a}$
independently controllable real parameters in the state $\left| \Psi
_{F}\right\rangle $. The state $\left| \Psi _{F}\right\rangle $ can be
written in general in the form
\begin{equation}
\left| \Psi _{F}\right\rangle =\sum_{n_{h}=0}^{N_{a}}b\left( n_{h}\right)
\left| N_{a},n_{h}\right\rangle ,  \label{psif}
\end{equation}
where the Dicke states $\left| N_{a},n_{h}\right\rangle $ are defined by Eq.
(\ref{dicke}), and the complex superposition coefficients $b\left(
n_{h}\right) $ are functions of $\theta _{m_{i}},\varphi _{m_{i}}$.
Superpositions of the Dicke states have $2N_{a}$ degrees of freedom, which
exactly equals to the number of control parameters $\theta _{m_{i}},\varphi
_{m_{i}}$.

Actually, we can prove that an {\it arbitrary superposition} of the Dicke
states $\left| N_{a},n_{h}\right\rangle $ (i.e., the state $\left| \Psi
_{F}\right\rangle $ with any coefficients $b\left( n_{h}\right) $) is
obtainable by choosing an appropriate set of control parameters $\theta
_{m_{i}},\varphi _{m_{i}}$. For the proof, we write the state (\ref{psif})
in the form $\left| \Psi _{F}\right\rangle =b\left( N_{a}\right) c\left(
N_{a}\right) \sum_{n_{h}=0}^{N_{a}}b^{\prime }\left( n_{h}\right) \left(
s_{0}^{\dagger }\right) ^{n_{h}}\left( s_{1}^{\dagger }\right)
^{N_{a}-n_{h}}\left| G\right\rangle $, where $b^{\prime }\left( n_{h}\right)
=c\left( n_{h}\right) b\left( n_{h}\right) /\left[ b\left( N_{a}\right)
c\left( N_{a}\right) \right] $, and without loss of generality we have
assumed $b\left( N_{a}\right) \neq 0$. Each of the atomic excitation
operators $P_{m_{i}\mu }$ can be expressed as $P_{m_{i}\mu }\propto \left(
s_{0}^{\dagger }-r_{m_{i}\mu }s_{1}^{\dagger }\right) $, where the complex
coefficient $r_{m_{i}\mu }$, determined by the real parameters $\theta
_{m_{i}},\varphi _{m_{i}}$, is the relevant control parameter. To prepare a
desired state $\left| \Psi _{F}\right\rangle $ with the superposition
coefficients $b^{\prime }\left( n_{h}\right) $, we need to choose the
parameters $r_{m_{i}\mu }$ to satisfy the algebraic equation $%
\prod_{i=1}^{N_{a}}\left( s_{0}^{\dagger }-r_{m_{i}\mu }s_{1}^{\dagger
}\right) =\sum_{n_{h}=0}^{N_{a}}b^{\prime }\left( n_{h}\right) \left(
s_{0}^{\dagger }\right) ^{n_{h}}\left( s_{1}^{\dagger }\right) ^{N_{a}-n_{h}}
$. It immediately follows from this equation that the parameters $%
r_{m_{i}\mu }$ should be the $N_{a}$ solutions of the $N_{a}$th-order
algebraic equation $\sum_{n_{h}=0}^{N_{a}}b^{\prime }\left( n_{h}\right)
x^{n_{h}}=0$, where $x$ denotes the variable. In the complex domain, there
always exist $N_{a}$ solutions to the $N_{a}$th-order algebraic equation,
and the parameters $r_{m_{i}\mu }$ are uniquely determined from these
solutions if we do not care about the order of the excitation operators $%
P_{m_{i}\mu }$ (note that they commute with each other).

This proves constructively that we can generate any superposition of the
Dicke states by choosing appropriate parameters $\theta _{m_{i}},\varphi
_{m_{i}}$. Of course, to prepare such a superposition, the success
probability of the scheme is typically significantly smaller than that for
preparation of a Dicke state. However, for a few atoms, it is still
reasonable to employ this scheme to prepare arbitrary superpositions of the
states $\left| N_{a},n_{h}\right\rangle $. Eq. (\ref{psif}) represents the
complete set of states in the symmetric subspace for all the atoms, and is
the largest set that can be prepared without separate addressing of
different atoms. The symmetric states $\left| \Psi _{F}\right\rangle $
include many interesting states as their special cases, such as the
well-known $N_{a}$-party GHZ\ state $\left| \Psi _{GHZ}\right\rangle =\left(
\left| N_{a},0\right\rangle +\left| 0,N_{a}\right\rangle \right) /\sqrt{2}$,
which is simply a superposition of two Dicke states.

In summary, we have presented a methodology for efficient
engineering of multi-atom entanglement in optical cavities. Our
schemes are inherently robust to many sources of practical noise
and technical imperfections, and thus well fit the status of the
current experimental technology. Our estimates suggest that it
would be reasonable to exploit these ideas to generate
high-fidelity entangled states for a sample of $N_{a}\sim 10$
atoms based on current capabilities in cavity QED \cite{12}, which
represents a very exciting experimental possibility.

{\bf Acknowledgments}: This work was supported by the Caltech MURI
Center DAAD19-00-1-0374, by the NSF Grants EIA-0086038 and
PHY-0140355, and by the Office of Naval Research. L.M.D. was also
supported by the CSF, the CAS, and the "97.3" project
2001CB309300.

\begin{figure}[tbp]
\epsfig{file=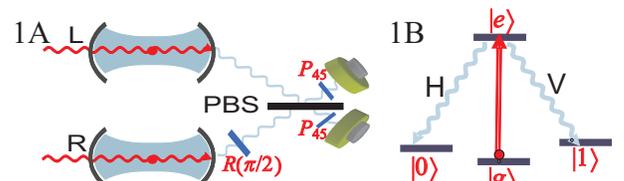,width=8cm} \caption{1a. The schematic setup
to generate entanglement between two atoms in different cavities
{\bf L} and {\bf R}. (2b). The relevant atomic level structure and
the laser configuration.}
\end{figure}

\begin{figure}[tbp]
\epsfig{file=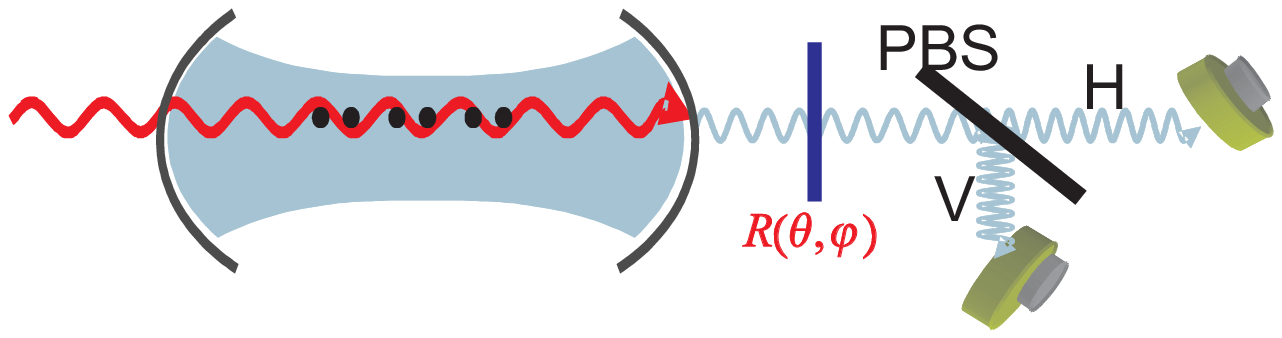,width=8cm} \caption{1a. The schematic setup
to generate entanglement between multiple
atoms in the same cavity. The polarization rotator $R\left( \protect\theta ,%
\protect\varphi \right) $ is only required for generation of
superpositions of the Dicke states.}
\end{figure}

\begin{references}
\bibitem{1}  N. D. Mermin, {\it Phys. Rev. Lett. }{\bf 65}, 1838 (1990).

\bibitem{2}  J. Bollinger, W. M. Itano, D. Wineland, D. Heinzen, {\it Phys.
Rev. A} {\bf 54}, 4649 (1996).

\bibitem{3}  M. A. Nielsen, I. L. Chuang, {\it Quantum Computation and
Quantum Information}, (Cambridge University Press, UK, 2000).

\bibitem{4}  C. Cabrillo, J. I. Cirac, P. G-Fernandez, P. Zoller, {\it Phys.
Rev. A} {\bf 59}, 1025 (1999).

\bibitem{5}  S. Bose, P. L. Knight, M. B. Plenio, V. Vedral, {\it Phys. Rev.
Lett.} {\bf 83}, 5158 (1999).

\bibitem{5'}  M. B. Plenio, S. F. Huelga, A. Beige, and P. L. Knight, {\it %
Phys. Rev. A} {\bf 59}, 2468 (1999).

\bibitem{5''}  J. Hong and H.-W. Lee, {\it Phys. Rev. Lett.} {\bf 89},
237901 (2002).

\bibitem{6}  A. S. Sorensen, K. Molmer, quant-ph/0206142

\bibitem{7}  I. E. Protsenko, G. Reymond, N. Schlosser, P. Grangier,
quant-ph/0206007.

\bibitem{8}  L. M. Duan, M. D. Lukin, J. I. Cirac, P,. Zoller, {\it Nature}
{\bf 414}, 413 (2001).

\bibitem{9}  L. M. Duan, {\it Phys. Rev. Lett.} {\bf 88}, 170402 (2002).

\bibitem{9'}  L. Mandel and E. Wolf, {\it Optical Coherence and Quantum
Optics}, Cambridge University Press (1995).

\bibitem{10}  W. D\"{u}r, G. Vidal and J. I. Cirac, {\it Phys. Rev. A} {\bf %
62}, 062314 (2000).

\bibitem{11}  P. Boyer and M.~A.~Kasevich, {\it Phys.~Rev.~A} {\bf 56},
R1083 (1997).

\bibitem{11'}  A. Cabello, {\it Phys. Rev. A} {\bf 65}, 032108 (2002)

\bibitem{12}  J. McKeever {\it et al.}, quant-ph/0211013.

\bibitem{12'}  A. Kuhn, M. Hennrich, and G. Rempe, {\it Phys. Rev. Lett.}
{\bf 89}, 067901 (2002).

\bibitem{12''}  Y. Shimizu {\it et al.}, {\it Phys. Rev. Lett.} 89, 233001
(2002).

\bibitem{13}  T. Pellizari {\it et al}., Phys. Rev. Lett. {\bf 75}, 3788
(1995).

\bibitem{13'}  J. I. Cirac {\it et al}., Phys. Rev. Lett. {\bf 78}, 3221
(1997).

\bibitem{14}  L.-M. Duan, A. Kuzmich, H.J. Kimble, quant-ph/0208051.

\bibitem{15}  D. F. Walls, and G. J. Milburn, {\it Quantum Optics},
Springer-Verlag (1994).

\bibitem{15a}  M. D. Lukin, S. F., Yelin, and M. Fleischhauer, {\it Phys.
Rev. Lett.} {\bf 84}, 4232-4235 (2000).

\bibitem{16}  J.-W. Pan {\it et al., Phys. Rev. Lett. }{\bf 86}, 4435 (2001).
\end{references}
\end{document}